\documentclass[article,aps,twocolumn,amsmath,amssymb,superscriptaddress,nofootinbib,english]{revtex4-1}
\usepackage{graphicx}% Include figure files
\usepackage{dcolumn}% Align table columns on decimal point
\usepackage{bm}% bold math
\usepackage{epsfig}
\usepackage{graphicx}
\usepackage{hyperref}
\usepackage[usenames]{color}
\usepackage{url}

\hypersetup{
    colorlinks=true,
    linkcolor=red,
    citecolor=blue,
}

\def\etal{{\frenchspacing\it et al.}}

\def\be{\begin{equation}}
\def\ee{\end{equation}}
\def\ba{\begin{eqnarray}}
\def\ea{\end{eqnarray}}
\begin{document}

\title{Fables of reconstruction: controlling bias in the dark energy equation of state} 

\author{Robert G. Crittenden}
\affiliation{Institute of Cosmology and Gravitation, University of Portsmouth, Portsmouth, PO1 3FX, UK}
\author{Gong-Bo Zhao}
\affiliation{Institute of Cosmology and Gravitation, University of Portsmouth, Portsmouth, PO1 3FX, UK}
\author{Levon Pogosian}
\affiliation{Department of Physics, Simon Fraser University, Burnaby, BC, V5A 1S6, Canada}
\author{Lado Samushia}
\affiliation{Institute of Cosmology and Gravitation, University of Portsmouth, Portsmouth, PO1 3FX, UK}
\affiliation{Abastumani Astrophysical Observatory, Ilia State University, 2A kazbegi Ave, Tbilisi, GE-0160, Georgia}
\author{Xinmin Zhang}
\affiliation{Theoretical Physics Division, Institute of High Energy Physics, Chinese Academy of Science, P.O.Box 918-4, Beijing 100049, P.R.China}
\affiliation{Theoretical Physics Center for Science Facilities (TPCSF), Chinese Academy of Science, Beijing 100049, P.R.China}

\begin{abstract}
We develop an efficient, non-parametric Bayesian method for reconstructing the time evolution of the dark energy equation of state $w(z)$ from observational data.  Of particular importance is the choice of prior, which must be chosen carefully to minimise variance and bias in the reconstruction.  Using a principal component analysis, we show how a correlated prior can be used to create a smooth reconstruction and also avoid bias in the mean behaviour of $w(z)$.  We test our method using Wiener reconstructions based on Fisher matrix projections, and also against more realistic MCMC analyses of simulated data sets for Planck and a future space-based dark energy mission. While the accuracy of our reconstruction depends on the smoothness of the assumed $w(z)$, the relative error for typical dark energy models is $\lesssim10\%$ out to redshift $z=1.5$.  
\end{abstract}

%\date{\today}

\maketitle

\section{Introduction}

The nature of dark energy (DE), the source driving the acceleration of the universe in the framework of general relativity, has remained a mystery since the acceleration was first discovered \cite{RiessPerl}. With the accumulating observational data, including supernovae (SN), cosmic microwave background radiation (CMB) and large scale structure (LSS) data, we hope to understand whether DE is a cosmological constant or whether it is dynamical.   Dynamical dark energy is characterised by having an equation-of-state $w \neq -1$, and which in general could be a function of redshift $w(z)$ or, equivalently, the scale factor $w(a)$.  One of the key goals in dark energy observations is to constrain the equation of state and its evolution.

The time evolution of $w$ can, in principle, be reconstructed from data using either parametric or non-parametric methods (e.g. \cite{Sahni:2006pa}). Non-parametric methods have the advantage that they do not assume an $ad~hoc$ functional form of $w(z)$, which could lead one to miss evidence for a different kind of evolution.   Unfortunately, non-parametric methods generically contain many degrees of freedom which can lead to parameter degeneracies, making them difficult to explore.  

Many efforts have been made in the literature to develop non-parametric methods, which can vary greatly depending on the nature of the data 
and the choices of the functions to be reconstructed.  One particular focus has been using SN and other standard candle data to reconstruct various ways of paramterizing the background cosmology; the techniques either fit directly to SN magnitudes or their luminosity distances $D_L(z)$, or to more indirect quantities such as dark energy density, $\rho_{DE}(z)$, the expansion history, $H(z)$, and the equation of state $w(z)$, or much more indirect quantities such as the potential of the DE scalar field(s) \cite{Huterer:1998qv,Saini:1999ba,Chiba:2000im,Huterer:2000mj,Weller:2001gf,Alam:2003sc,Daly:2003iy,Shafieloo:2005nd,Dick:2006ev,Shafieloo:2007cs,Clarkson:2010bm}.

Unfortunately, many of these techniques are specific to luminosity distance measurements, or more generally to measures of the background expansion rate, making it difficult to include other types of data.  The background measurements can be limited by our knowledge of other parameters, in particular the value of the dark energy density today, $\Omega_{DE}$ \cite{Chiba:2000im,Alam:2003sc}.  Thus, it is useful to include measurements which could be sensitive to the growth of structure; such measurements, including weak lensing, galaxy clustering and the integrated Sachs-Wolfe effect (ISW), can help to break the degeneracies and can potentially be the basis for consistency tests of the DE paradigm \cite{Mortonson:2008qy, Mortonson:2010fy}. 

When comparing background and structure evolution data, it is useful to focus on more indirect parameterisations like the equation of state, $w(z)$.  Non-parametric approaches typically would expand $w(z)$ in terms of bins or basis functions.  A powerful tool is the principal component approach, which uses forecasts of the future experiments in the form of Fisher matrix projections to find linear combinations of these basis functions which are independent and well determined \cite{ Huterer:2002hy, Crittenden:2005wj, Albrecht:2007qy, Tang:2008hm, Albrecht:2009ct, Kitching:2009yr, Mortonson:2010px,Clarkson:2010bm}.   

One reconstruction approach is to make a truncation of the best constrained principal components, but the number of modes kept is an open question.  This often is decided without reference to the data, potentially missing evidence because it is not expected.   This simple truncation can lead to significant biases and unrealistically small errors in the reconstruction where the data are poor or are absent entirely \cite{ Huterer:2002hy}.    It is clear that some bias in reconstructions is inevitable, particularly  when one attempts to reconstruct $w(z)$ in regimes where the data are poor or are absent entirely.  However, it is worth investigating whether a better reconstruction method can be found.    

In a fully Bayesian approach, one can explicitly specify a prior on the behaviour of $w(z)$, which then determines which modes are kept in the reconstruction, based on where the evidence from the data outweighs the prior.  But the crucial question is then to understand the best way to specify priors in order to minimise the bias, while at the same time reducing the impact of data noise in the reconstruction.  
Here we extend previous work, where we assumed that $w(z)$ could be treated as a Gaussian random field with a proposed correlation function \cite{Crittenden:2005wj}; this correlation function effectively enforces a smoothness on the reconstruction.   Our approach is close in spirit to recently proposed Gaussian Process (GP) method \cite{Holsclaw:2010nb,Holsclaw:2010sk}, which applied a similar prior to reconstructions of $w(z)$ using SN data, though there are important differences as described below.

In this paper we discuss how to choose a prior and implement it in a computationally efficient method to reconstruct $w(z)$ non-parametrically from the observed data. As we will show, it is very accurate (with relative error $\lesssim$ 10\% for a range of dark energy models), computationally cheap and fast, straightforward to implement, and can be used to analyse any kind of cosmological data.
After discussing general issues regarding the choice of prior correlation functions, we test our reconstruction accuracy using the simulated mock data for Planck \cite{Planck} and a future space-based mission.  We then conclude with a discussion of how to extend this method to other parameterization and how to calculate more realistic priors.  

\section{Reconstruction methods}

Reconstruction methods in general have been well explored; here we focus on a Bayesian method, so to make our assumptions as explicit as possible. Whatever the method, one needs some metric to evaluate the quality of the reconstructions.  One natural choice of metric, or risk function, is the mean squared error (MSE) of the binned equation of state, between the true model and the reconstructed one:
\begin{equation}
\label{eq:mse1}
{\rm MSE} \equiv \sum_i (w^{\rm true}_i - w^{\rm recon}_i)^2. 
\end{equation}
The MSE is not the only possible choice, but it is common and we adopt it here.  

Our Bayesian method must assume an \textit{a priori} probability distribution in the space of models.  In the absence of data, the reconstruction will return \textit{some} fiducial model, ${\bf w}^{\rm fid}$, which is the peak of the prior distribution; if the fiducial model differs from the true one, the mean reconstructed model will be biased, where ${\bf w}^{\rm mean} \equiv \langle{\bf w}^{\rm recon} \rangle $ is averaged over the ensemble of possible data consistent with the true model.  
The ensemble average of the MSE can be shown to be the sum of two terms: the variance with which the mean model will be reconstructed and the bias between the true model and the mean one:
\begin{equation}
\label{eq:mse2}
\langle {\rm MSE} \rangle = \sum_i   \langle (w^{\rm mean}_i - w^{\rm recon}_i)^2 \rangle +  (w^{\rm true}_i - w^{\rm mean}_i)^2. 
\end{equation} 
The challenge of reconstruction is to keep both types of errors small; a stronger prior will reduce the variance of the reconstruction, but will increase the bias if the fiducial and true models do not match.  

Here we focus on a fully Bayesian reconstruction method.  In it, we simply assume a prior probability distribution on the function we wish to reconstruct, and our reconstructed function is the maximum of the posterior probability distribution, which is the product of the data likelihood and the prior.  Thus, the reconstruction method is entirely specified by the definition of the prior and this makes our reconstruction assumptions fully explicit.    

We try to choose the prior to minimise the bias; however, this is intrinsically subjective because it depends on the expected `true' model.  In principle, this should be determined by theoretical considerations and/or previous observations.  Given that we are making projections based on the most optimal future data, previous observations are not expected to have significant impact, making theoretical considerations key.  Theoretical priors can be made by considering ensembles of possible models, either using a combination of parametric models for the equation of state, or by using physical models; e.g., by putting a prior on the space of possible quintessence potentials.  However, it is clear that this choice could vary wildly between theorists.   

One of the problems of using a non-parametric models is the sensitivity of answers to the binning scheme assumed.  If few bins are assumed, then the answers are well determined, but the resulting reconstruction has unphysical discrete structures resulting from the binning.  But if many bins are assumed, some degrees of freedom (d.o.f.) will not be constrained by the data, resulting in large variance and slow convergence of Monte Carlo Markov chain (MCMC) methods.  An advantage of the Bayesian approach is that we can use many bins, but then constrain the residual degrees of freedom by imposing a prior on the space of functions.  This prior tends to make the functions smoother, reflecting our preconceptions of how the equation of state should evolve.   

Another common approach to reconstructions is to simply smooth the data, e.g. by using some implementation of a low-pass filter, on the assumption that the models will not have high frequency variations.  This equates to an infinitely strong prior, which does not allow high frequency modes in the reconstruction, no matter how strong the evidence for them may be.  While this is possible to implement with the methods we describe, it seems more reasonable to set some strong but finite prior based on theoretical considerations; that way, high frequency modes could, in principle, enter the reconstruction if the evidence for them becomes sufficiently strong (where the definition of sufficient is based purely on theoretical grounds.)                 

Here we focus on phenomenological choices for the prior, and make recommendations based on simple considerations, but allow that this will be up to individual choice.

\section{The correlated prior}

\subsection{The correlation function}

The prior distribution could be an arbitrarily complex multi-variate probability distribution for the non-parametric amplitudes.  
We assume this distribution is Gaussian, meaning it is specified by a covariance matrix describing fluctuations around a mean or fiducial 
model, ${\bf w}_{\rm fid}$.   Here for simplicity, we further assume that the covariance matrix is a function only of the scale factor and in particular that it is translation independent, depending only on the difference in scale factor $|a - a'|$.

These assumptions allow us to specify the prior using a simple one-dimensional function, defined in \cite{Crittenden:2005wj} as 
\begin{equation}
 \xi_w (|a - a'|) \equiv \left\langle [w (a) - w^{\rm fid}(a)][w(a') - w^{\rm fid}(a')] \right\rangle,
\end{equation}
which can be used to generate the likelihood for any $w(a)$.  This has the effect of reducing the degrees of freedom of the function, effectively binding together neighboring bins below some specified correlation length $a_c.$  This is precisely the kind of prior assumed in the Gaussian process approach \cite{Holsclaw:2010nb,Holsclaw:2010sk}, though the shape and parametrization of the correlation function can differ considerably. In Ref. \cite{Crittenden:2005wj}, we applied this smoothness prior to the Fisher forecasts, and found it a natural way to quantify how much we could learn from future data.

Here we choose the scale factor, $a$, as our independent variable, which is somewhat arbitrary and in our earlier work  \cite{Crittenden:2005wj} instead used redshift.  A function which is translation invariant in one variable will not be precisely translation invariant in another, meaning the priors cannot be equivalent.  However, if our answers strongly depend on such differences, then likely the priors have been made too strong.    

\subsection{Implementation of the correlation prior }
\label{sec:mcmc}

We start by discretising $w(a)$ at $a \geq a_{\rm min}$ using bins uniform in scale factor $a$, and use a wide bin for $a < a_{\rm min}$. We have checked that using uniform bins in ln$(a)$ or $z$ returns a consistent result when the number of bins $N$ is large enough, namely, $N\geq 20$ with $a_{\rm min}=0.4$, which we adopt in this work. Note that we use {\tt tanh} bins rather than tophat bins so that the time derivative of $w$ is stable at any $a$, which is an important issue for calculating the dark energy perturbations when using CMB and LSS data. 

Given some correlation shape and choice of fiducial model, it is straight forward to discretize them given our choice of binning. Let us assume the i$^{th}$ bin is from $a_i$ to $a_{i} + \Delta$, and for simplicity we will assume that all bins have the same width $\Delta = a_{i+1} - a_i$.  The equation of state averaged over each bin is given by
\begin{equation}
w_i = \frac{1}{\Delta} \int_{a_i}^{a_i+\Delta} da \, w(a).
\end{equation}
We can write the variation from the true model averaged over the bin as, $\delta w_i =  w^{\rm true}_i - w^{\rm fid}_i .$
Calculating the covariance matrix of the binned equation of state is then straightforward:
\be
C_{ij}\equiv \langle \delta w_i \delta w_j  \rangle = {1 \over \Delta^2} \int_{a_i}^{a_i+\Delta} da \int_{a_j}^{a_j+\Delta} da' \xi_w (|a - a'|) .
\label{eq:corr-matrix}
\ee
Given this, the prior for the model can be written as 
\begin{equation}
{\cal P}_{\rm prior}({\bf w}) \propto  e^{-({\bf {w}} -{\bf {w}}^{\rm fid})^T {\bf {C}}^{-1}({\bf {w}} - {\bf {w}}^{\rm fid} )/2} .\label{eq:prior-dist}
\end{equation} 
Subsequently, in the MCMC, we minimize the total posterior $\chi^2$ defined as
\be
\chi^2=\chi^2_{\rm data} + \chi^2_{\rm prior} \ ,
\label{eq:chi2}
\ee
where
\be
\chi^2_{\rm prior}=-2\ln {\cal P}_{\rm prior} = ({\bf {w}} -{\bf {w}}^{\rm fid})^T {\bf {C}}^{-1}({\bf {w}} - {\bf {w}}^{\rm fid}).
\ee

In Sec.~\ref{sec:exploring} we show that one can marginalize over possible choices of  ${\bf w}^{\rm fid}$, or define a procedure by which ${\bf w}^{\rm fid}$ is defined using an averaging of ${\bf {w}}$. In such a case, the
prior probability can be written as
\begin{equation}
{\cal P}({\bf w}) \propto  e^{-{\bf w}^T {\tilde {\bf C}}^{-1}{\bf w} /2} , 
\label{eq:Pprior}
\end{equation} 
where $\tilde{\bf C}$ is a modified version of the correlation matrix, meaning that 
\be
\chi^2_{\rm prior}= {\bf w}^T {\tilde {\bf C}}^{-1}{\bf w} \ .
\ee

The correlation function typically provides a frequency dependent prior: high frequency oscillations are suppressed by the prior, while the low frequency modes are largely unaffected and are so dependent on the data. Providing
a prior stabilizes the high frequency variances and allows us to focus on the more interesting low frequency modes. It also significantly improves the MCMC convergence, as the flat directions are now curtailed by the prior. Also, as long as there are sufficient bins compared to the correlation length, the prior largely wipes out dependence on the precise choice of binning.

\subsection{The Wiener filter} \label{sec:Wiener}

To reconstruct $w(a)$ from real data, we simply fold the theoretical prior into our MCMC search using Eq.~(\ref{eq:chi2}) and search for the optimal model. One can gain analytical insight into reconstructions using the correlated prior by exploiting similar arguments used for  \textit{Wiener filtering}, a well understood methodology for reconstructing a signal in the presence of noise.  The Wiener approach can be applied in this context once we recognise that the Fisher matrix (or its inverse) describes the expected noise covariance in the $w(a)$ reconstruction, while our theoretical prior describes our expected signal covariance.  

We assume that we have some noisy data vector which is the best fit to the observed data, and we wish to estimate the best actual $w(a)$ given the data has noise described by the Fisher matrix, $F_{ij}$. 
The expected likelihood is given by 
\begin{equation}
{\cal P}({\bf w}^{\rm obs} | {\bf w}) \propto e^{-({\bf w}^{\rm obs} -{\bf w} )^T {\bf {F}}({\bf w}^{\rm obs} -{\bf w})/2}. 
\label{eq:like-Fisher}
\end{equation}
Combining this with the theoretical prior defined above, we find the posterior distribution 
${\cal P}({\bf w} | {\bf w}^{\rm obs})$.  
We define the reconstructed equation of state as that which maximizes the posterior; we find 
\begin{equation}
{\bf F}({\bf w}^{\rm recon} -{\bf w}^{\rm obs}) + {\bf C}^{-1} ({\bf w}^{\rm recon} - {\bf w}^{\rm fid}) =0, 
\end{equation}
which has the solution, 
\begin{equation}
{\bf w}^{\rm recon} = {\bf F}^{-1}({\bf C} + {\bf F}^{-1})^{-1}{\bf w}^{\rm fid} + {\bf C}({\bf C} + {\bf F}^{-1})^{-1} {\bf w}^{\rm obs}.  
\label{eq:Wiener}
\end{equation}
Assuming that the noise dominates in the higher frequencies, this is effectively a low-pass filtering of the data combined with a high-pass filtering of the fiducial model.  Given most choices of fiducial models will be smooth, the latter should have minimal effect.  

As mentioned above and detailed in Sec.~\ref{sec:exploring}, it is possible to eliminate the explicit dependence on ${\bf w}^{\rm fid}$ by introducing a modified correlation matrix $\tilde{\bf C}$ with the prior probability given by Eq.~(\ref{eq:Pprior}). Then the maximum posterior solution becomes simply
\begin{equation}
{\bf w}^{\rm recon} =  \tilde{\bf C}(\tilde{\bf C} + {\bf F}^{-1})^{-1} {\bf w}^{\rm obs}.
\label{eq:Wiener-tilde}
\end{equation}

This analytic solution is useful for projections, which we present in Sec.~\ref{sec:projections}. When working with real data, it is very difficult to estimate ${\bf w}^{\rm obs}$, which optimises the observations alone, because flat directions in parameter space make the MCMC convergence impossible when the number of bins is large. But when searching for ${\bf w}^{\rm recon}$, such flat directions are curtailed by the prior.  

\section{Exploring specific correlated priors}
\label{sec:exploring}

\subsection{The choice of correlation function}

The choice of amplitude and shape of the prior can critically effect the reconstructed $w(a)$ and other conclusions, such as the number of dark energy parameters which can be constrained.  The prior should incorporate both our theoretical prejudices and other earlier cosmological data which have not been explicitly included in the present analysis.  Theoretically, one might examine the equation of state dynamics of the quintessence field, beginning with some measure on the space of quintessence potentials (e.g. \cite{Huterer:2006mv}, also see \cite{Crittenden:2007yy}.)  Earlier data can also be used to decide on the prior.  However, it is not clear how useful this is, as the earlier data will naturally be weaker than the data being considered, so it would be surprising if it resulted in stronger constraints on any degree of freedom.  If the experiment provides independent constraints, it should be incorporated explicitly rather than using a heuristic prior model which can misrepresent the distribution of the information.    
Here we attempt a more pragmatic approach, by using simulated data and attempting to reconstruct a number of `typical' dark energy models.  

One choice is to assume that the prior is diagonal in the binning, effectively assuming the correlation is a delta function \cite{Albrecht:2009ct}.  If all bins have the same prior, the correlation matrix is proportional to the identity matrix, which has the advantage that the principal components of data remain unchanged when the prior is included.  However, such a prior lacks the bin independence and smoothing properties of a more realistic prior choice.  

In previous work \cite{Crittenden:2005wj}, we worked with redshift $z$ as the variable, and modeled the correlation function using a form $\xi_w (\delta z) =  \xi_w (0) /[1 + (\delta z/z_c)^2]$. Here, we adopt the same form expressed in terms of the scale factor:
\be
\xi_w (\delta a) =  { \xi_w (0) \over 1 + (\delta a/a_c)^2} \ ,
\ee 
where $a_c$ describes the typical smoothing distance, and $\xi_w(0)$ which is a normalising factor that relates to the amplitude of the expected variance of $w(a)$. In what follows, we will refer to this model as CPZ. The normalisation can also be chosen by specifying the allowed variance of the average equation of state, 
\begin{equation}
\sigma^2_{\bar{w}} \equiv \int_{a_{\rm min}}^{1} \int_{a_{\rm min}}^{1} {da \ da' \ \xi_w (a -a') \over (1-a_{\rm min})^2} 
 \simeq \frac{\pi \xi(0) a_c}{1-a_{\rm min}}.
\end{equation}
The precise shape assumed was somewhat arbitrary, chosen for simplicity and ease of use; while the correlation is expected to decrease, the decline could be steeper a steeper power, or exponential rather than a power law.  The Gaussian process \cite{Holsclaw:2010nb,Holsclaw:2010sk} work instead typically assumes an exponential fall off, with `hyper-parameters' determining the shape; some distribution is assumed for these hyper-parameters and then they are marginalised over.           

\subsection{Comparing different correlation shapes}

\begin{figure}[tb]
\includegraphics[scale=0.33]{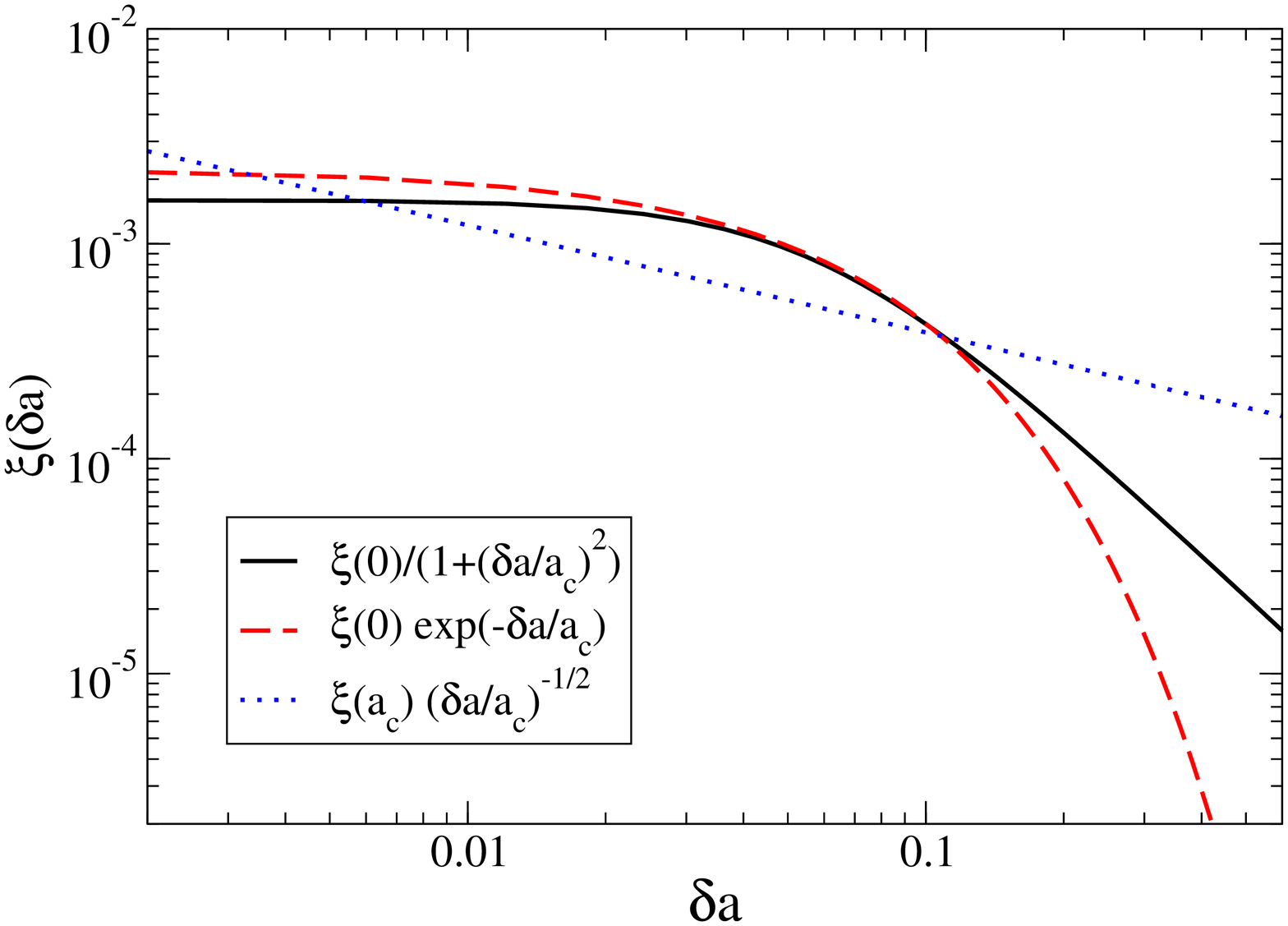}
\caption{Different prior shapes normalised to the same mean variance and $a_c = 0.06$: CPZ (black, solid), exponential (red, dashed), power law (blue, dotted).  
}\label{fig:corr_shape}
\end{figure}

\begin{figure}[tb]
\includegraphics[scale=0.33]{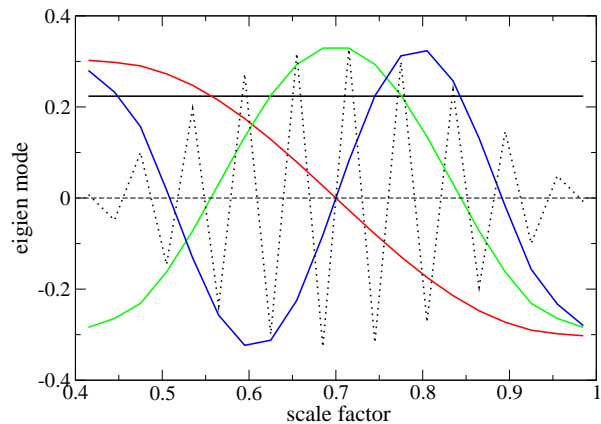}
\caption{Some of the normalised eigenvalues of the prior.
Changes to the prior description have little effect on the modes themselves.  The modes are largely ordered in frequency (black, red, green, blue), with the lowest frequencies the least constrained by the prior.  The highest frequency mode (dashed) is the most constrained.  The deviation from the simple Fourier transform is due to the boundaries on $w$.     
}\label{fig:evec-prior}
\end{figure}

We compare a few different shapes for the correlation functions, to see the impact of their qualitative features.  In addition to the form discussed above, $\xi_{\rm CPZ} (\delta a) =  \xi_w (0) /[1 + (\delta a/a_c)^2],$ we also consider an exponential fall off, $\xi_{\rm exp} (\delta a) =  \xi_w (0) e^{-\delta a/a_c},$ and a general power law form, $\xi_{\rm pow} (\delta a) =  (\delta a/a_c)^{-n}.$  We normalise each to the same mean variance, $\sigma^2_{\bar{w}}$ and set $a_c=0.06$, apart from the power law form where specifying the variance fixes the value of $a_c$.  We then compare the discretised correlation matrices over the range $a = [0.4, 1.]$ The different shapes are compared in Figure \ref{fig:corr_shape}.

We first consider the behaviour of the functions at small separations, where it can either approach a constant or diverge for the power law models.  If the slope is too steep, the discretised diagonal correlation also diverges unless a small scale cut-off is imposed.  In the limit that this cut-off becomes small, the correlation matrix effectively becomes diagonal which we consider as a separate case.  Below, we focus on the power law case only for $n = 1/2,$ a shape shallow enough to avoid the divergence.  

The impact of these different ways of defining priors is perhaps most clearly seen in the eigenvalues of the associated prior matrices.  To illustrate this, we choose a binning (20 bins between $a= 0.4$ and $a=1.$) and calculate the corresponding correlation matrices, and find the eigenmodes and their eigenvalues in this binning.  

Perhaps surprisingly, virtually all the ways of defining the prior result in similar prior eigenmodes, and in identical ordering for the eigenvalues of these modes.  A few of the modes are shown in Fig. \ref{fig:evec-prior}, where it is clear that the modes correspond closely with the Fourier basis, apart from minor differences which arise from the scale factor boundaries.  Because correlations fall off with distance, the most constrained modes are those with the highest frequencies.  Slowly varying modes are least constrained by the prior.  (In the limit of a diagonal correlation matrix, the eigenvalues are degenerate and the mode definitions are arbitrary.)       

\begin{figure}[tb]
\includegraphics[scale=0.33]{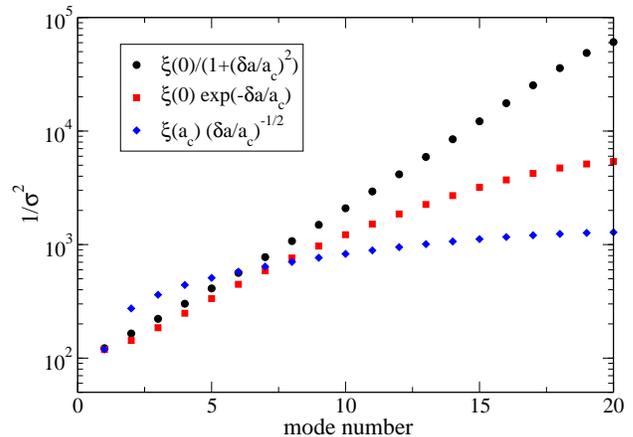}
\caption{Different prior eigenvalues normalised to the same mean variance: CPZ (black, circles), exponential (red, squares), power law (blue, diamonds).  The normalisation results in the same lowest eigenvalue, and modes increase in frequency.  
}\label{fig:eval_prior}
\end{figure}

However, the different correlation shapes significantly change the associated eigenvalues, as can be seen in Fig. \ref{fig:eval_prior}.   Because the models are normalised to the same mean variance, the lowest eigenvalues are the same.  However, the spacing of the higher values is significantly different.  For the CPZ model, the eigenvalues are nearly equally spaced in log, and are steepest of these three models for a given $a_c$, leading to the most constrained high frequency modes.  This spacing directly depends on $a_c$: the longer the correlation distance, the steeper the spectrum.  In the limit of small $a_c$, the matrix becomes diagonal.  Because of its relatively simple behaviour and transparent dependence on its parameters, we adopt the CPZ form as our default model below.

\subsection{Changing the fiducial model}\label{sec:fid}

The prior distribution function, as defined in (\ref{eq:prior-dist}), depends explicitly on the choice of the fiducial equation of state $w^{\rm fid}(a)$.  For simplicity, we could assume the fiducial model to be $w=-1$, but arguably this is {\it ad hoc}.  Formally the priors should be chosen from fundamental physics;  a Gaussian prior centred at $w=-1$ is unphysical in many simple models that do not allow the so called `phantom' \cite{Caldwell:1999ew} region, $w <-1$, which violates the null energy condition.  Dynamical models like quintessence may asymptote to cosmological constant behaviour at early or late times (thawing or freezing models), but are always constrained to have $w \ge -1$. 
While phantom models are possible, models which cross the `phantom divide' (dubbed `quintom' models \cite{Feng:2004ad}) can also be a struggle to build in a natural way.  In our present phenomenological approach, it makes more sense to choose a prior which does not favour a particular value. Since we wish to test $\Lambda$CDM, we prefer to use a prior which does not favor $w=-1$ over another model, e.g., $w=-0.9$. 

One way to eliminate the dependence on $w^{\rm fid} (a)$ is to marginalise over it. For a constant $w^{\rm fid}$ this can be done analytically; we can write the fiducial model as ${\bf w}^{\rm fid} = w^{\rm fid} {\bf u}$, where ${\bf u} = (1, 1, 1, ...)$ and we assume a weak Gaussian prior on $w^{\rm fid}$ with variance given by $\sigma^2_{\rm fid}$. 
Marginalising over $ w^{\rm fid}$ has the effect of changing the effective correlation matrix: 
%\begin{eqnarray}
%\int d w^{\rm fid}  e^{-({w}^{\rm true}_i -{w}^{\rm fid}u_i) {C}_{ij}^{-1}({w}^{\rm true}_j -{w}^{\rm fid}u_j )/2 - (w^{\rm fid})^2/2\sigma^2_{\rm fid} } \nonumber \\ \propto %
%e^{-{w}^{\rm true}_i  \tilde{C}_{ij}^{-1}{w}^{\rm true}_j/2}
%\end{eqnarray}
\begin{eqnarray}
{\cal P}_{\rm prior}&\propto& \int d w^{\rm fid} e^{-({\bf w} -{w}^{\rm fid}{\bf u}) {\bf C}^{-1}({\bf w} -{w}^{\rm fid}{\bf u} )/2 - (w^{\rm fid})^2/2\sigma^2_{\rm fid} } \nonumber \\ 
& \propto &  e^{-{\bf w}^T {\tilde {\bf C}}^{-1}{\bf w} /2} \ ,
\end{eqnarray}
where 
\begin{equation}
\tilde{C}_{ij}^{-1} =  {C}_{ij}^{-1} - \frac{{C}_{ik}^{-1} u_k u_l  {C}_{lj}^{-1}}{\sigma^{-2}_{\rm fid} + u_l {C}_{lk}^{-1} u_k} \ .
\end{equation}
Such expressions are a common result when marginalising over nuisance parameters (e.g. \cite{Bond:2001bd}) and can be simply inverted using the Sherman-Morrison formula \cite{sherman-morrison}, a special case of the Woodbury formula \cite{woodbury}, to find:
\begin{equation}
\tilde{C}_{ij} = {C}_{ij} + \sigma^{2}_{\rm fid} u_i u_j. 
\label{eq:ctilde-marg}
\end{equation}  
Here we see the utility of introducing a prior on the fiducial value, however weak -- a flat prior corresponds to an infinite variance, which makes $\tilde{C}$ undefined.  Most importantly, marginalising assigns large noise to the flat mode in the prior and ensures that for reconstructions this mode is determined by the data alone.

\begin{figure}[tb]
\includegraphics[scale=0.33]{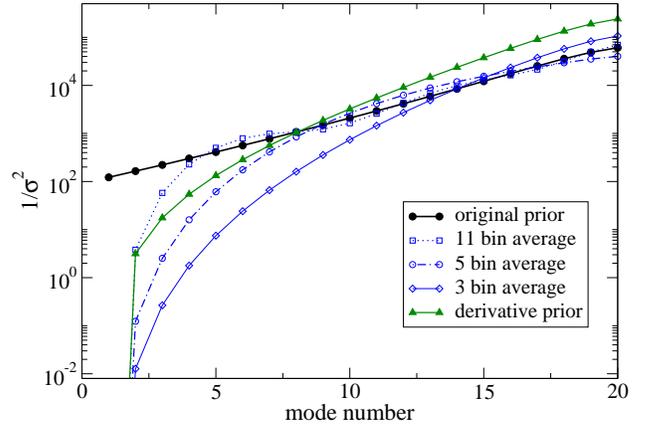}
\caption{ Impact on the eigenspectrum of various transforms considered.  The original CPZ spectrum is in black (solid circles).  Marginalising over the mean only zeroes out the first eigenvalue.  Using a running average tends to reduce the lowest frequency modes while leaving the higher frequency constraints unchanged. (11 bin average - open circles, 5 bin average - open squares, 3 bin average - open diamonds.)  Taking the CPZ prior form for the derivative results in the green triangles; the spectrum steepens and the first mode is zeroed.   
}\label{fig:eval-local}
\end{figure}

An alternative approach is to define the fiducial model to be a function of the assumed equation of state (${\bf w}^{\rm fid}({\bf w})$) rather than being fixed; for example, it could float to a constant value given by the average of the equation of state, 
\begin{equation}
{w}^{\rm fid} = \bar{w} \equiv \frac{1}{a_f - a_i} \int_{a_i}^{a_f} da \, w(a) = \frac{1}{N} \Sigma_i w_i .
\end{equation}
Such a change, where the fiducial model is a linear transformation of the underlying model, ${\bf w}^{\rm fid} = S {\bf w}$ is also equivalent to simply changing the effective correlation matrix.  The explicit dependence of the prior on the fiducial model 
is absorbed into a modified correlation matrix, 
\begin{equation}
\tilde{C}_{ij}^{-1} = (\delta_{ik} - S^T_{ik}){C}_{kl}^{-1}  (\delta_{lj} - S_{lj}). 
\end{equation} 
In the case above where the fiducial model is the average over all bins, we have $S_{ij} =u_i u_j/N$, implying 
\begin{equation}
\tilde{C}_{ij}^{-1} = {C}_{ij}^{-1}  - \frac{1}{N} (u_i u_k{C}_{kj}^{-1} + {C}_{ik}^{-1} u_k u_j) + \frac{1}{N^2} u_i u_j u_k {C}_{kl}^{-1} u_l. 
\label{eq:ctilde-global}
\end{equation}
This is similar to the result for marginalisation and, like that case, can result in a degenerate direction unless an additional prior is assumed; without this, the matrix cannot be inverted.

The implementation of a `floating' fiducial model can be based on other smoothing schemes. We might instead define the fiducial model as a more local average of the true model, which reduces the prior penalty for the longer wavelength modes, making them more responsive to the data.  For example, we could have
\be 
w_{i}^{\rm fid}=\sum_{|a_j-a_i|\leq a_c}w^{\rm true}_j/N_j
\label{eq:local}
\ee 
where $N_j$ denotes the number of the neighboring $w$ bins around the $i$th bin $w_i$ within the `correlation' radius $a_c$.  We tried other smoothing methods, such as a Gaussian kernel scheme, and found very similar results.

These modifications to the fiducial model significantly affect the eigenvalues, particularly the one for the constant $w$ mode, which is effectively reset to zero (or the prior that it has been explicitly given.)   Both approaches, marginalising over the mean value or redefining the fiducial model to the global average, do this while leaving the other eigenvalues unchanged, just as expected.   Using a more local definition of the fiducial model also reduces the constraints on other long wavelength modes while leaving the shorter ones mostly unchanged, as can be seen in Fig. \ref{fig:eval-local}. We adopt this local averaging method of Eq.~(\ref{eq:local}) when performing the reconstructions in Sec.~\ref{sec:data}

\subsection{Correlations of $w(a)$ derivatives }

Another way to ensure that the average of the fiducial model does not impact  the reconstruction is to apply a prior instead to the derivative of the $w(a)$ function.  Essentially, we specify a correlation matrix for the difference of the equation of state between bins, 
\begin{equation}
\langle (w_i - w_{i+1})(w_j - w_{j+1}) \rangle = D_{ij}.
\end{equation}  
This has one less dimension than the correlation of the $w$ bins itself.  Again, we can define a Gaussian prior based on this correlation matrix.

\begin{equation}
{\cal P}({\bf w}) \propto  e^{-({w}_i -{w}_{i+1}) ^T {D}_{ij}^{-1}({w}_j -{w}_{j+1} )/2} .
\end{equation} 
This can be thought of as a prior on the bin values themselves with an alternative inverse correlation defined by, 
\begin{equation}
\tilde{C}_{ij}^{-1} = {D}_{ij}^{-1} - {D}_{i-1,j}^{-1} - {D}_{i,j-1}^{-1} + {D}_{i-1,j-1}^{-1}
\label{eq:ctilde-derivs}
\end{equation}
The indices of $\tilde{C}$ span $[1,N]$, while those for $D$ span $[1,N-1]$; where an index for $D$ exceeds its range (i.e. $i=0, N$), the inverse matrix is taken to be zero. 

 While $\tilde{C}_{ij}^{-1}$ is well defined in this way, it is not invertible because the constant $w$ mode is not constrained in any way.  If required, this can be combined with a weak prior on the constant $w$ mode to yield a correlation matrix which is invertible.  

Assuming the fiducial CPZ correlation for the derivative of $w$ (rather than for $w$ itself) results in a new spectrum of eigenvalues.  As expected, the constraint on the mean mode is reset to zero.  However, it also results in a steeper spectrum than the original one, as can be seen in Fig. \ref{fig:eval-local}.  This is roughly equivalent to using a smaller correlation length in the original prescription, and so would not qualitative change the analysis below.  

\section{Reconstructing the equation of state }
\label{sec:data}

\begin{figure}[tb]
\includegraphics[scale=0.33]{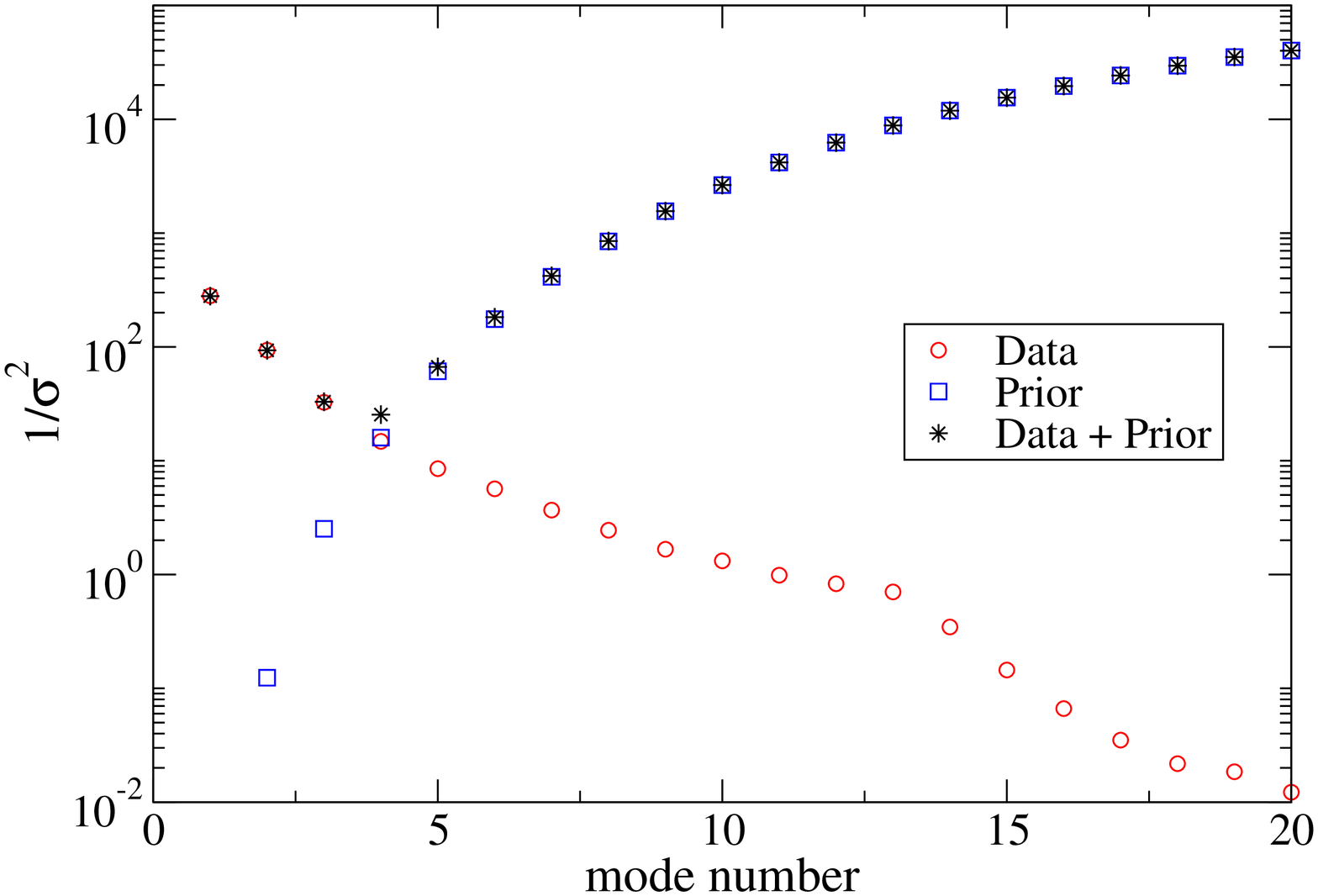}
\caption{Eigenvalues from the data (red circles) and prior independently (blue squares) and combined (black stars).   The modes are arranged roughly from low frequency to high frequency, though this correspondence is not exact when the data are included.  The low frequency modes are constrained by the data, while the high frequency modes are constrained by the prior.     
}\label{fig:eval-tot}
\end{figure}

Next we combine the priors we have defined with the data to see how effective they will be in reconstructing $w(a)$. We do this in two ways: first, using calculations of the Fisher matrices for a set of observations to test Wiener reconstructions of some sample $w(a)$ functions, effectively assuming a simple Gaussian approximation to the likelihood.  Second, we create more realistic simulations of the expected CMB, SN and $H(z)$ data and reconstruct $w(a)$ by finding the maximum posterior model using MCMC techniques.  

\subsection{Data assumptions}

We test our reconstruction algorithms by combining the CMB distance prior \cite{Komatsu:2010fb} from Planck \cite{Planck} with supernovae (SN) and galaxy-galaxy correlation measurements potentially possible for a future space-based dark energy mission.  Note that our assumptions are close to those that could be achieved from a next generation survey such as Euclid, but do not exactly match those now expected for the default Euclid mission \cite{Euclid, Euclid-official}.  For the CMB, we use a Fisher matrix forecast based on the Planck data. For the SN data, we assume 4000 SN distributed in 14 redshift bins from $z=0.15$ to $z=1.55$ for a deep SN survey \cite{Astier:2010qf}. We also include 300 low-$z$ SN from the Nearby Supernova Factory \cite{WoodVasey:2004pj} to improve the constraint and assume a Gaussian noise with a variance $\sigma=0.13$ for all the data points. 

The measurement of the expansion rate at different redshifts $H(z)$ can tighten constraints on $w(a)$, since the DE energy density is directly related to the Hubble rate at a given redshift (e.g., \cite{Stern:2009ep}). One of the most promising ways of estimating $H(z)$ from future spectroscopic surveys is to observe the Alcock-Paczynski effect in the power spectrum or correlation function of galaxies \cite{Kaiser:1987qv}, exploiting features such as the baryon acoustic oscillations (BAO) and the power spectrum peak. The BAO feature in the correlation function is at a fixed comoving scale which is determined by the physics at recombination; measuring the redshift difference of this feature in the radial correlation function provides a direct local measurement of $H(z)$.  We divide the galaxies  ($0.35<z<2.5$) into bins of width $\Delta z = 0.1$ and predict the measurement errors $\sigma_{H(z)}$ in each redshift bin, using the Fisher matrix technique of \cite{Samushia:2010ki}. We assume that the spaced-based survey would cover 20,000 ${\rm deg}^2$ surveying area and will detect emission line galaxies with the redshift precision of $\sigma_{z}=0.001(1+z)$. For our projections, we adopt the values derived in \cite{Geach:2009tm} for the number density of emission line galaxies and the bias model of \citep{Orsi:2010mj}. 

Given these data models, we consider the following set of cosmological parameters,
\be
{\bf P} \equiv (\omega_{b}, \omega_{c},
\Theta_{s}, \tau, n_s, A_s, w_1, ..., w_{20},\mathcal{N}),
\label{eq:paratriz} 
\ee
where $\omega_{b}$ and $\omega_{c}$ are the physical baryon and cold dark matter densities relative to the critical density respectively, $\Theta_{s}$ is the ratio of the sound horizon to the angular diameter distance at decoupling, $\tau$ denotes the optical depth to reionization, $n_s$ and $A_s$ are the primordial power spectral index and amplitude respectively and $w_1,...,w_{20}$ denote the 20 $w$ bins uniform in $a$. We also include and marginalize over one nuisance parameter $\mathcal{N}$ for supernovae, which accounts for the calibration uncertainty in measuring the supernova intrinsic luminosity. The details of the Fisher forecast and the related survey parameters can be found in \cite{Pogosian:2005ez} and \cite{Zhao:2008bn}.  After marginalising over the cosmological and nuisance parameters, the projected constraint on a constant equation of state model is $\sigma_{\bar{w}} = 0.015.$ % Figure of Merit.   

\subsection{Reconstruction Forecasts}

A Fisher matrix analysis provides a means of projecting the constraints possible for the experiments, under the optimistic assumption that the resulting likelihood is Gaussian (Eq. \ref{eq:like-Fisher}.)  Combined with our assumed Gaussian prior, we can analytically forecast the expected reconstructions when assuming a particular underlying $w(a)$ using the Wiener filter approach described in Section \ref{sec:Wiener}.  This approach also allows us to quantify the expected bias and variance for a given input model.  

\subsubsection{Combined eigenmodes}

When combining data with the prior, the effective resulting Fisher matrix changes from ${\bf F} \rightarrow {\bf F} + \tilde{\bf C}^{-1}$. 
Many of the general aspects of reconstructions can be understood by considering the eigenvalues of this combined matrix.  Roughly speaking, the data constrain the low frequency modes the best, while the prior constrains the high frequencies, and the most constrained modes of each survive in the combined matrix.  However, this is only approximate since the data and prior have different eigenmodes; in particular, the data modes are more weighted to the lower redshifts.

\begin{figure}[tb]
\includegraphics[scale=0.45]{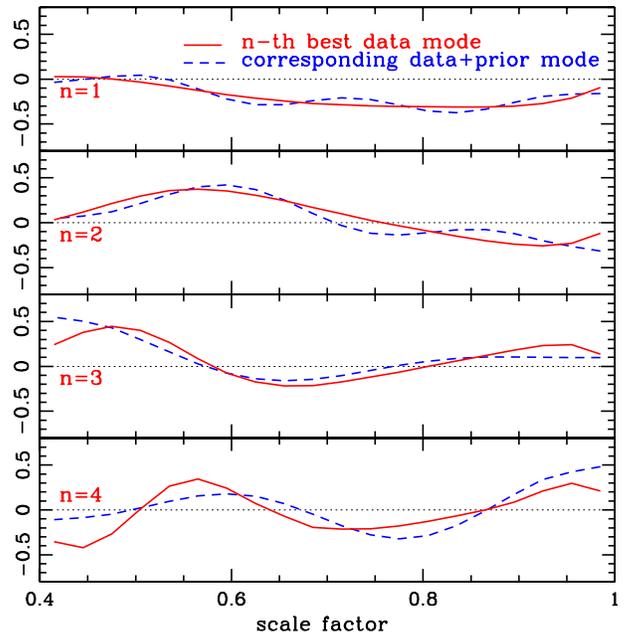}
\caption{The first four best constrained eigenmodes of the data covariance matrix (solid lines), along with the corresponding modes of the data+prior covariance (dashed lines). As seen in Fig.~\ref{fig:eval-tot}, the data eigenvalues become comparable to the prior eigenvalues at the $n=4$ mode, and the $n > 4$ modes are completely determined by the prior.
}\label{fig:emodes-tot}
\end{figure}

Fig. \ref{fig:eval-tot} shows the impact of combining the data and prior on the eigenvalues.  The cross over point is crucial, as it determines the reconstruction bias, while the slope of the prior eigenvalues will determine the variance of the reconstructions.  With the data described above and our fiducial prior choice, only 4 data modes survive to play a role in the reconstructions. The surviving modes are shown in Fig.~\ref{fig:emodes-tot}. One can see that, while the shapes are distorted by the prior, their basic features remain in tact.

More modes would survive if the data were better, or if the prior were weaker (e.g. reducing the prior normalisation $\sigma_{\bar{w}}$ or the slope $a_c$.)   The benefit of a weaker prior is that more modes would be used to reconstruct the data, giving smaller potential bias.  The cost would be significantly more variance in the reconstructions.  The key issue is whether the true underlying $w(a)$ will require more modes to achieve a small bias.  

\begin{figure*}[t]
\includegraphics[scale=0.16]{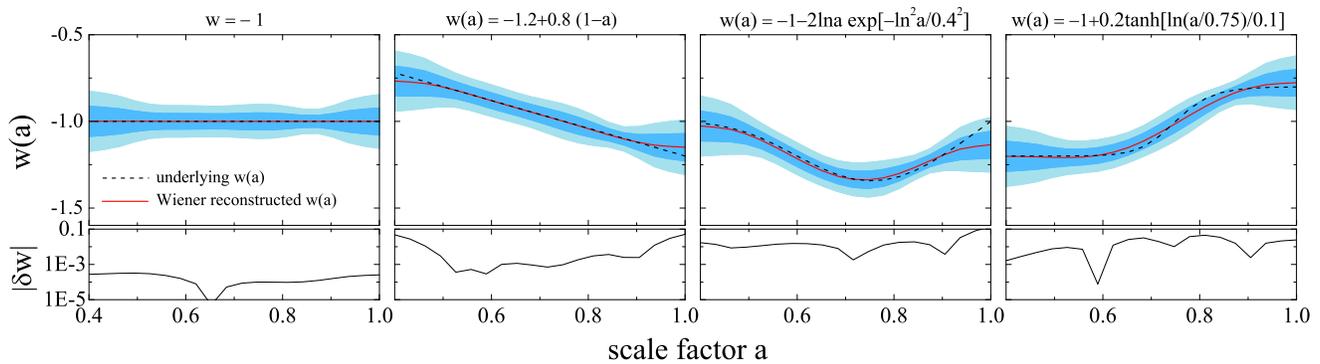}
\caption{Wiener reconstructions of the various test functions, based on the Fisher projections for a joint data set of SN + $H(z)$ + CMB distance prior (Planck), and using our fiducial prior assumptions.       
}\label{fig:recon}
\end{figure*}

\subsubsection{Wiener filter projections}
\label{sec:projections}
If the Gaussian approximation to the likelihoods are reasonable, then the Fisher matrix should allow a good approximation to the expected errors on the measured $w_i$ bins.  We can make realisations of the observed binned $w_i$ data by assuming some true underlying model and adding to it random noise derived using the principal components of the Fisher matrix.  Each principal component is assigned a random Gaussian amplitude with the appropriate variance, and these are summed to get the total noise in the binned data.  These realisations are then Wiener smoothed using Eq. (\ref{eq:Wiener-tilde}) to get the reconstructed $w(a)$.  
(Note that when $\tilde{\bf C}$ makes negligible constraint on the average, then the term 
%in Eq. (\ref{eq:Wiener}) dependent on the fiducial model,  
${\bf F}^{-1}(\tilde{\bf C} + {\bf F}^{-1})^{-1}{\bf w}^{\rm fid}$ can be neglected if the fiducial model itself is constant, making Eqns. (\ref{eq:Wiener}) and  (\ref{eq:Wiener-tilde}) equivalent.)  

With these assumptions, we can derive analytically the mean reconstructed models, as well as the variance and bias expected.  
For a given ${\bf w}^{\rm true}$, the average reconstructed model is 
\begin{equation}
{\bf w}^{\rm mean} =  \tilde{\bf C}(\tilde{\bf C} + {\bf F}^{-1})^{-1} {\bf w}^{\rm true}.
\end{equation}
This implies that that bias contribution to the MSE is  
\begin{equation}
 \sum_i  (w^{\rm true}_i - w^{\rm mean}_i)^2 = || {\bf F}^{-1}(\tilde{\bf C} + {\bf F}^{-1})^{-1} {\bf w}^{\rm true} ||^2 .
\end{equation} 
We can similarly predict the variance contribution under these assumptions to find, 
\begin{equation}
\sum_i  \langle (w^{\rm mean}_i - w^{\rm recon}_i)^2 \rangle  = {\rm Tr}[({\bf F} + \tilde{\bf C}^{-1})^{-2}{\bf F}]. 
\end{equation} 
Note that in this approximation, the variance is independent of the assumed model.  We have confirmed these analytic results using 
20,000 realisations for each input model.    

We consider four basic underlying phenomenological models (${\bf w}^{\rm true}$) to test the ability of our algorithm to capture different kinds of evolving features in $w(a)$: 
\begin{eqnarray}\label{eq:4models}
w_{\rm const} = & -1.0 \nonumber \\
w_{\rm lin} = & -1.2+0.8(1-a)\nonumber \\
w_{\rm feat} = & -1+2{\rm ln}a\cdot{\rm exp}(-{\rm ln}^2a/0.4^2)\nonumber \\
w_{\rm trans} = & -1 + 0.2 \tanh(\ln(a/a_t)/\Delta)  \ .
\end{eqnarray}
We refer to these as the constant $w$ model, the linear model \cite{CPL}, the feature model (see discussions in \cite{feature}) and the transition model (e.g. \cite{Corasaniti:2002vg}), respectively.
For this section, we also consider a fifth `thawing' model based on the parametrization of \cite{Crittenden:2007yy} for a thawing quintessence which smoothly goes from $w=-1$ at high redshifts to $w=-0.8$ today.   The transition model changes from $w= -1.2$ at high redshifts to $w=-0.8$ at low redshifts, and we chose the transition to occur at $a_t=0.75$ with width of $\Delta =0.1.$

These functions and their reconstructions are shown in Fig. \ref{fig:recon}.  For these reconstructions we use our fiducial prior, which we define as the CPZ prior ($\sigma_{\bar{w}} = 0.02, a_c = 0.06 $) with a fiducial model as the average of bins within $\Delta a = 0.06$ given by Eq.~(\ref{eq:local}) (effectively the five bin average shown in Fig. \ref{fig:eval-local}.)   Note that $\sigma_{\bar{w}}$ primarily is used to normalise the high frequency priors; in fact, the local averaging means that effectively no prior is placed on $\bar{w}$.

One can see that on average the reconstructions are fairly unbiased, with practically no bias in the case of a constant $w$.  This is expected, as the smoother functions are well represented by the modes which survive the prior.  Models with stronger evolution, such as the feature model and the fast transition model, have the most bias.  However, even for these, the average MSE is dominated by the variance for this choice of prior.  

The variance can be decreased by strengthening the prior, but this effectively allows fewer data modes to survive and so increases the potential bias of the reconstructions.  Thus the MSE cannot be dramatically reduced except in those cases where the bias is intrinsically small.  For example, the constant $w$ modes are unbiased in this prescription, and so the variance can be made very small.  

The MSE, and its contributions from bias and variance for each input model are shown in Table \ref{tab:mse}.  One can see that the MSE figures are dominated by the variance, with the bias term being comparable in the transition model when the change is very quick.  This is intentional; a stronger prior does not significantly reduce the MSE for most of the models we assume (the exception being the constant model), and given the choice it seems better to have a slightly noisier reconstruction than one which could be significantly biased.   

\begin{figure}
\vspace{0.2in}
\includegraphics[scale=0.33]{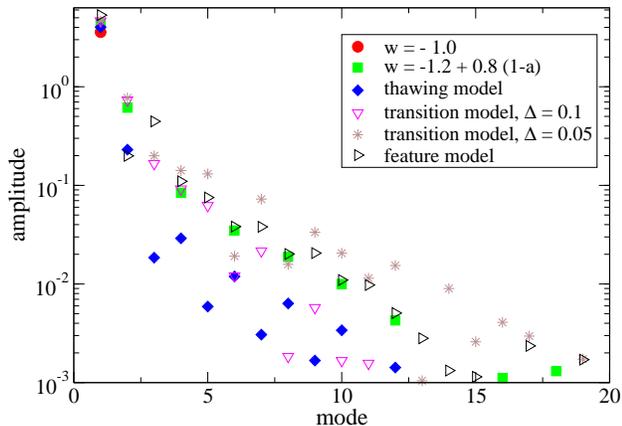}
\caption{Amplitudes of the various test functions expanded in the prior eigenmodes.   Those functions which are best reconstructed are those dominated by the low mode numbers.   Those with higher frequency features are reconstructed with more bias.      
}\label{fig:amp-target}
\end{figure}

The relative biases can be easily understood by looking at the input models expanded in the eigenmodes of the prior, as shown in Fig. \ref{fig:amp-target}.  We see that the models which are reconstructed with least bias (constant, thawing model, slow transition model) have little support beyond the fifth prior mode, while the others have some higher frequency components which are wiped out by the prior.

\begin{figure*}[t]
\includegraphics[scale=0.16]{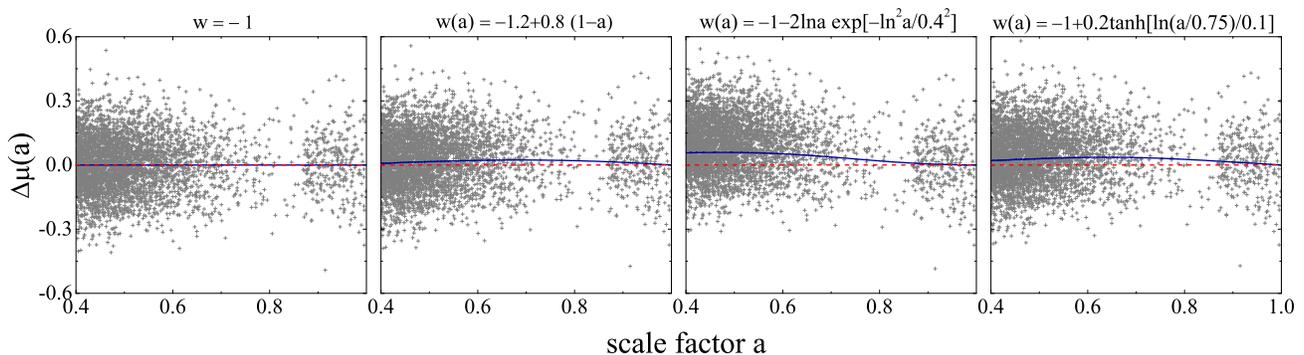}
\caption{The simulated SN magnitudes for a future deep survey as described in \cite{Astier:2010qf} for four phenomenological dark energy models listed in Eq (\ref{eq:4models}). The blue solid line shows the relative SN magnitude with respect to that of the $\Lambda$CDM model. The red dashed line shows $\Delta\mu=0$ to guide eyes.      
}\label{fig:mock}
\end{figure*}

\begin{figure*}[t]
\includegraphics[scale=0.16]{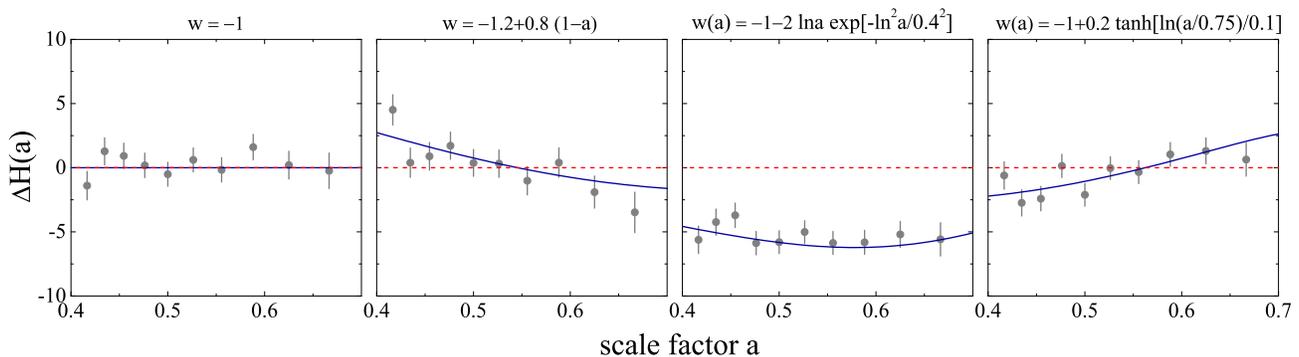}
\caption{The simulated $H(a)$ measurements for a future half-sky survey as described in \cite{Samushia:2010ki} for four phenomenological dark energy models listed in Eq (\ref{eq:4models}). Again, the blue solid line shows the difference in $H(a)$ with respect to that of the $\Lambda$CDM model. The red dashed line shows $\Delta H(a) = 0$ to guide eyes.      
}\label{fig:mock}
\end{figure*}

\begin{table}[tb]
\begin{tabular}{|c|c|c|c|}
\hline
Model    &   MSE   &   Bias   &   Variance    \\
\hline
$w = -1$       	&  0.064 	& -  	& 0.064   \\
Linear			&  0.068  	& 0.004 & 0.064   \\
Feature 		&  0.081    & 0.017	& 0.064  \\
Transition ($\Delta = 0.1$) 	&  0.071	& 0.007 & 0.064  \\
Transition ($\Delta = 0.05$) 	&  0.095	& 0.031 & 0.064  \\
Thawing 	   	&  0.065	& 0.001	& 0.064  \\
\hline 

\end{tabular}
\caption{A comparison of MSE results for different assumed models, broken down into bias and variance contributions. }
\label{tab:mse}
\end{table}

\subsection{Reconstructions from simulated data}

\begin{figure*}[tbp]
\includegraphics[scale=0.15]{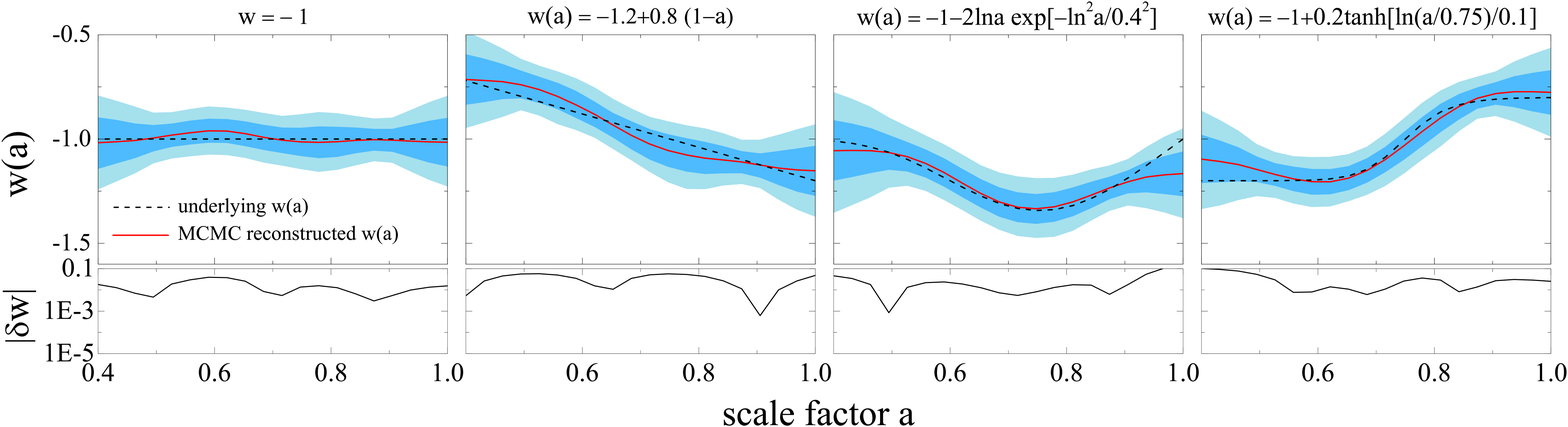}
\caption{The reconstructed $w(a)$ from a joint mock datasets of SN + $H(z)$ + CMB distance prior (Planck). We show the best fit model (red solid), the true model we put in (black dashed) with 68 and 95\% CL. error (dark and light shaded bands) in the upper panels, and the absolute value of the relative difference in the lower panels. Unlike the Wiener results, these are based on a single realisation of the data. 
}\label{fig:SN_deep}
\end{figure*}

% Simulated data sets 
We next explore how our reconstruction method works for more realistic data, because Fisher forecasts tend to be optimistic and ignore potential non-Gaussianities in the likelihood.  Here we look at single realisations of the potential data sets based on the models in Eq.~(\ref{eq:4models}).  For these we then perform a blind MCMC reconstruction, adding to the data $ \chi^2_{\rm data}$ a similar term for the prior  $\chi^2_{\rm prior} $ as described in Sec~\ref{sec:mcmc}.

To generate the simulated data, we set the fiducial values of other cosmological parameters to values favoured by the WMAP7 observations \cite{Komatsu:2010fb}, i.~e. $\Omega_m=0.263$ and $H_0=71$ km/s/Mpc. We then simulate the SN luminosity data by randomly generating  the SN redshifts and errors, while respecting their expected redshift density.  The simulated SN distance modulus $\mu$ of four phenomenological dark energy models (offset by that of the $\Lambda$CDM model) are shown in Fig. \ref{fig:mock}.  We similarly randomly generate $H(z)$ measurements in the redshift bins described above consistent with their expected errors. 

For the CMB data, we generate realisations for the three CMB distance parameters \cite{Komatsu:2010fb} consistent with their expected covariances expected from Planck. Note that for the real data analysis, one should use the full CMB spectrum data, which is also sensitive to dark energy through the integrated Sachs-Wolfe effect.  For such calculations, one must include dark energy perturbations, otherwise the derived constraints on cosmological parameters will be biased \cite{Weller:2003hw,Bean:2003fb,wmap3}. Furthermore, one should treat dark energy perturbations self-consistently when w crosses $w=-1$ \cite{Zhao:2005vj,Fang:2008sn}.  One must be careful even with the CMB distance prior, which can be biased if the model is significantly different from the $\Lambda$CDM for which it has been derived \cite{Li:2008cj}.

We simultaneously fit the 20 bins with $\Omega_m$ and $H_0$ to our combined mock dataset using a modified version of CAMB/CosmoMC \cite{Lewis:2002ah}, with the smoothness prior implemented. In Fig. \ref{fig:SN_deep} we show the reconstructed $w(a)$ from a joint mock dataset of SN + $H(z)$ +CMB distance prior (Planck). As shown, our algorithm can successfully capture all the DE models we hide in the data, and the reconstruction is very accurate -- the absolute value of the relative difference $\lesssim$ 10\% in all cases. 
The calculation of the prior $\chi^2_{\rm prior}$ adds virtually no overhead to the likelihood calculations, making the method quite efficient. 

It is important to emphasise that Figs. \ref{fig:recon} and \ref{fig:SN_deep}, while apparently very similar, are actually showing very different things.  Fig. \ref{fig:recon} shows the average reconstructed $w(a)$ over the ensemble of data realisations consistent with the experiments, and the errors show the distribution of the reconstructed $w(a)$ for these different data realisations.   Fig.  \ref{fig:SN_deep} instead shows the reconstructed $w(a)$ for a single simulated data set (similar to what we expect to get from real observations) and its error bars represent the uncertainty in the determination of $w(a)$ for this realisation.  This best fit differs from the input model both because of the average bias seen in Fig. \ref{fig:recon} and because of the randomness of the realised data.  However, the error regions are similar because they are related to very similar $\Delta \chi^2$ functions.

\section{Discussion}

Our reconstructions are very good, but some bias remains; this primarily takes the form of smoothing out the more quickly varying $w(a)$ behaviours.  While this can be made smaller by weakening the prior, the cost would be to increase the variance in the reconstruction.  The bias simply reflects our assumption that smoother $w(a)$ models are more likely.    
 
This is not an intrinsic limitation of the method, but reflects our choice to assume that $w(a)$ is smooth.  While we do not have to make this assumption, we do require some theoretical basis where smoothness (or a measure) can be assumed, such as the DE scalar field potential $V(\phi)$.  However, this is essentially the requirement that the given model is well enough posed to be falsifiable.  Any model description should allow for the calculation of the prior in some effective basis, such as bins of $w(a)$.

We could either perform the reconstruction directly using the parameters where the behaviour is expected to be smooth, or we could reconstruct in a more phenomenological basis, like $w(a)$, but use a prior in this basis derived assuming smooth behaviour in another basis.  Direct reconstruction of something like the scalar field potential is better in the sense that one keeps in contact with the theoretical context in which the prior makes sense; however, it makes it harder to compare the conclusions arising from different assumptions.     
Alternatively, we can derive a prior on $w(a)$, either arising from more fundamental physics or purely phenomenological models by calculating  the resulting $w$ correlation matrices $C$ by marginalising over whatever parameters appear in the description (e.g. \cite{Huterer:2006mv}).  

Inevitably, the simplifying assumptions we have made above for the form of these correlations will not hold when realistic models are considered. In particular, we assumed that the prior is Gaussian, and that the correlations are a function of $\delta a$ and are translation independent.   The latter is certainly not expected,  for example `freezing' models of quintessence will have larger variance at high redshifts, while `thawing' models will have larger variance at low redshifts.  Even translating the simple linear parameterisation, $w(a) = w_0 + w_a(1-a)$ with Gaussian priors on $w_0$ and $w_a$, into a correlation matrix on $w(a)$ turns out not to be translation invariant; the resulting variance is smallest at $a=1$, and increases with $1-a$. 

It is possible that including many different classes of models together, such as thawing and freezing models, might homogenise the correlations to some extent.  Alternatively, one could hope that such deviations of the covariance from our simplifying assumptions will not be significant enough to greatly impact the reconstructions.  However, this requires further exploration and we will investigate this in future work.

\section{Conclusions}

Here we have outlined a Bayesian method for reconstructing $w(a)$, which essentially involves multiplying the likelihood of a non-parametric description of $w(a)$ by a suitably defined prior.  This prior is defined in such a way as to make the reconstructions smooth and independent of the binning choice, eliminating flat directions in parameter space which are a generic problem for non-parametric approaches.  

With the right choice of prior, we can also eliminate the dependence on the fiducial model;  reconstructions where the data are poor or non-existent are instead determined by the assumed smoothness of the function, tending towards the mean of the measured $w(a).$   However, application of a prior always implies some bias; in this case, the bias is largest for models which are quickly changing in redshift or scale factor.   

The advantages of this approach are that it makes one's priors explicit and it is easy to implement; once one decides on the prior parameters, e.g. the correlation length and the strength of the prior, it is straightforward and fast to evaluate the prior likelihood for whatever choice of binning is used.  This can be combined with the data in an MCMC analysis, searching for the highest posterior solution for the $w$ bin amplitudes; indeed, by breaking the parameter degeneracies, the prior will speed up MCMC analyses.  

With the accumulating high-quality data of SN, CMB and LSS, our method is an ideal tool to help reveal the nature of dark energy in the near future.  Here we have focused on the methodology only, but in future work we will apply this prior to present data.  

%\acknowledgments

\acknowledgments We thank Dragan Huterer, Hubert Lampeitl, Michael Mortonson, Bob Nichol and Will Percival for useful comments and discussions.
RC and GBZ are supported by STFC grant ST/H002774/1,  LP by an NSERC Discovery Grant and LS by the European Research Council. XZ is supported in part by the National Natural Science Foundation of China under Grant Nos. 11033005, 10803001, 10975142 and also the 973 program No. 2010CB833000.  Numerical computations were done on the Sciama High Performance Compute (HPC) cluster which is supported by the ICG, SEPNet and the University of Portsmouth.

\end{document}